\documentstyle[12pt,benaip,mosttext,nodisplaylabel,epsf_jane,figures,generaldefs,mathdefs]{article}

\newcommand{\phimax}{\phi_{\small max}}
\newcommand{\bmax}{b_{\small max}}
\newcommand{\Po}{P_0}

\begin{document}

\bibliographystyle{benaip}


\renewcommand{\thepage}{}


\titleben{Morphology Selection of Nanoparticle Dispersions by Polymer Media}
%

\author{
JAEUP U. KIM\ $^{1}$ \ and \  BEN O'SHAUGHNESSY\ $^2$ \\ 
}

\maketitle

\ \\ 
{\large $^1$ Department of Physics, Columbia University, New York, NY 10027} \\
\ \\
{\large $^2$ Department of  Chemical Engineering, Columbia University, New York, NY 10027} \\
\
\\
Submitted To: {\large \em Physical Review Letters}


\pagenumbering{arabic}

\section*{ABSTRACT}

Designable media can control properties of nanocomposite materials by
spatially organizing nanoparticles. Here we theoretically study
particle organization by ultrathin polymer films of grafted chains
(``brushes''). Polymer-soluble nanoparticles smaller than a
brush-determined threshold disperse in the film to a depth scaling
inversely with particle volume. In the polymer-insoluble case,
aggregation is directed: provided particles are non-wetting at the
film surface, the brush stabilizes the dispersion and selects its
final morphology of giant elongated aggregates with a brush-selected
width.

\vii

PACS numbers:
  \begin{benlistdefault}
    \item [82.35.Np]
(Nanoparticles in polymers)
    \item [68.08.-p]
(Liquid-solid interfaces: Wetting)
     \item [82.35.Jk]
(Copolymers, phase transitions, structure)
\end{benlistdefault}

\pagebreak


The prospect of a new generation of materials and devices based on the
assembly of nanoparticles into spatially extended 2D and 3D
arrangements is a major driving force in the rapidly emerging field of
nanoscale research \citeben{chan:nanoporus_prl,fogg:cluster_prl}.
Nanoparticles, the ``designer molecules'' governing the macroscopic
behavior of these novel materials, can be constructed according to a
vast range of design principles, promising unprecedented tuning of
material properties.  Both for technology and for fundamental research
in condensed matter physics, the implications are far-reaching.

This letter addresses a major challenge in the field: how to spatially
organize the nanoparticles.  The interparticle {\em medium} plays a
crucial role in this regard, and one would like to ``design'' media to
achieve different spatial arrangements leading to different
interparticle couplings and entirely different macroscopic properties.
Much recent research has examined polymeric media spontaneously
forming nanostructured phases as templates for particle organization
\citeben{chan:nanoporus_prl,fogg:cluster_prl,hamdoun:composite_prl,%
balazs:numeric_prl}.  Moreover, polymer-particle composites are often
conveniently processable with important implications for thin film and
other technologies
\citeben{fogg:cluster_prl,liu:nanoletters_prl,zehner:gold_prl}.

In this letter we develop a systematic theory of ultrathin polymer
films as organizing media to achieve 2D nanoparticle arrangements.
The latter are central to information storage media applications and
of fundamental importance to 2D electronic phenomena.  We are directly
motivated by recent studies
\citeben{fogg:cluster_prl,liu:nanoletters_prl} of metallic and
semiconducting nanoparticles in films of end-tethered polymer chains
forming ``polymer brushes,'' and in diblock copolymer materials whose
lamellae offer a similarly 2D brush-like environment
\citeben{hamdoun:composite_prl,zehner:gold_prl}.  A major objective is
that this work offer guidance to such experiments by identifying the
key physical variables to achieve nanoparticle dispersions and control
morphology.

\begin{strip_ft}
{0}
{0}{7}{figure1}
{0}
{0}{7}{\caption[Propagation]{
\renewcommand{\baselinestretch}{1} \footnotesize \bf
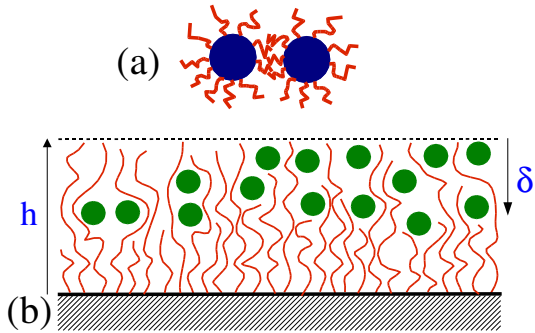
(a) Typical nanoparticles comprise inorganic cores of several $nm$,
plus stabilizing oligomeric coronas for compatibilization with polymer
medium.  (b) Schematic of polymer film containing nanoparticles in
contact with air. Chains are grafted to substrate, forming a stretched
``brush'' of height $h$ at large grafting density $\sigma$.
}
\label{nanopols_scheme}}
\end{strip_ft}

Given a nanoparticle-containing film of substrate-grafted polymer
chains of $N$ units at grafting density $\sigma$
(fig. \ref{nanopols_scheme} (b)) in contact with air, we will see there
are two distinct classes of behavior depending on whether the
particles are (A) soluble or (B) insoluble in the free polymer medium
(ungrafted chains).  Our principle conclusions are as follows:

({\bf A}) For {\bf polymer-soluble} particles (well-compatibilized)
their size $b$ is crucial: (i) Sufficiently small particles disperse
freely within the polymer brush film.  (ii) Above a threshold size
$b^*$, equilibrium particle penetration is limited to a depth $\delta$
and the film has a loading capacity $\phimax$.  Both $\delta$ and the
maximum particle density $\phimax$ scale inversely with particle
volume $b^3$. (iii) Particles larger than a second threshold
$\bmax \approx (N/\sigma)^{1/4}$ cannot penetrate the film.

({\bf B}) {\bf Polymer-insoluble} particles tend to aggregate in the
polymer film, of course.  This aggregation is {\em directed} by the
polymer brush: (i) The brush imposes a severe free energy penalty for
aggregate growth in two lateral directions simultaneously (parallel to
substrate): large {\em anisotropic} aggregates tend to form, elongated
in one lateral direction.  (ii) As growth continues, aggregates are
expelled towards the brush-air interface and the final state depends
on the polymer-nanoparticle-air interfacial tensions.  If the
nanoparticles tend to wet the polymer film surface, they will spread
to form a near-monolayer.  In the non-wetting case, growth is
ultimately arrested and the final morphology consists in large
elongated aggregates, embedded in the film, whose width is selected by
the brush.

\begin{strip_ft}
{0}
{0}{7}{figure2}
{0}
{0}{7}{\caption[Propagation]{
\renewcommand{\baselinestretch}{1} \footnotesize \bf
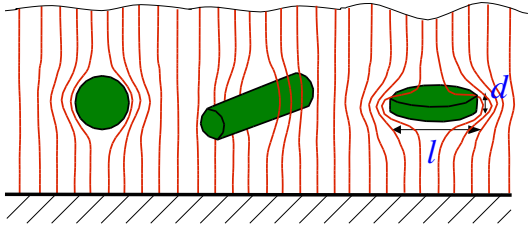
The degree an inclusion disrupts a polymer
brush depends strongly on shape. To within prefactors, the free energy
costs $\Delta F$ for a cylinder and a sphere of the same volume are
equal because chain distortion is determined by cylinder radius rather
than its length. But chain configurations are far more strongly
distorted by a disk and $\Delta F$ is much higher, increased by the
disk aspect ratio $l/d$.
}
\label{aggregate_shape}}
\end{strip_ft}

Before proceeding, a few remarks concerning these conclusions may be
helpful.  Inevitably, polymer-insoluble nanoparticles will eventually
phase separate if thermodynamic equilibrium is accessible.  Thus the
experimental procedure must not only forcibly introduce the particles
into the polymer film, but thereafter their eventual expulsion and
phase separation must be prevented by physical constraint.
(Alternatively, kinetics may naturally intercede by inhibiting
particle fluidization in the air medium.)  Our conclusions for case
(B) have assumed this, and this turns out to be the most interesting
case: provided the particles are non-wetting on the film surface, the
brush {\em stabilizes} the nanocomposite and {\em selects the
morphology} of giant anisotropic nanoparticle aggregates with a
selected lateral dimension.  The physical origin of these effects is
that aggregates extended simultaneously in two lateral directions
impose radical distortions on the brush chain configurations (see
fig. \ref{aggregate_shape}), whereas chains are relatively undistorted
by an aggregate elongated in only one direction, around which they can
``flow'' by taking the shortest route.  This leads to a much lower
free energy cost.

For the remainder of this letter we briefly outline our calculations
leading to these conclusions; we then compare with experiment, and we
finish by proposing optimal future experimental strategies based on
this work.


In typical grafted polymer films, chain end anchoring is achieved by
attaching an end block chemically similar to the
substrate \citeben{liu:nanoletters_prl}.  It is well known that as grafting
density increases brush height $h$ increases (see
fig. \ref{nanopols_scheme} (b)) due to incompressibility, $h=\sigma
N$, and chains are forced to stretch away from the substrate.  Since a
Gaussian chain has stretching energy $(1/2)\ h^2/N$
\citeben{gennes:book}, this leads to a brush stretching energy density
or ``pressure'' $P_0 = (1/2) (h/N)^2 = (1/2)\sigma^2$.  (Units are
chosen so both $kT$ and the size of one monomer unit are unity, so the
volume of one chain is $N$.)

Now suppose nanoparticles are forced into such a brush (see
fig. \ref{nanopols_scheme} (a)); typically these are
surface-stabilized metallic, metal oxide or semiconductor
nanocrystals \citeben{fogg:cluster_prl,hamdoun:composite_prl,zehner:gold_prl}.
We first consider polymer-soluble particles, case (A).  To determine
nanoparticle density distributions $\phi_{nano}({\bf r})$ within the
brush, our starting point is the following brush free energy as a
function of chain configurations ${\bf R}(n,{\bf \rho})$ ($n$ is the
monomer label, $0\le n \le N$, ${\bf r}\equiv (x,y,z)$ with $z$
orthogonal to substrate):
\begin{eq}{fart}
F  = \int d{\bf\rho} g({\bf\rho}) \frac{3}{2}
                \int_0^N dn \left(\frac{\partial{\bf R}}{\partial n}\right)^2 
				+ \int d{\bf r} P({\bf r})
	        \left(\phi_{pol}({\bf r})+\phi_{nano}({\bf r})-1\right) +
                F_{nano}[\phi_{nano}]
\end{eq}
where $\phi_{pol}({\bf r}) \equiv \int d{\bf\rho} g({\bf\rho})
\int_0^N dn {\bf \delta}({\bf r}-{\bf R}(n,{\bf \rho}))$ is the
polymer density and $g(\rho)$ is the distribution of chain end
locations ${\bf \rho}$, ${\bf R}(0,{\bf \rho}) \equiv {\bf \rho}$.
The first term represents chain stretching free energy
\citeben{semenov:semenov_prl}, the pressure-like field $P({\bf r})$
enforces incompressibility and $F_{nano}$ describes interactions
involving nanoparticles.

Now if one deletes the nanoparticle terms, this gives the free energy
considered by Semenov for the pure non-solvated brush
\citeben{semenov:semenov_prl}.  Semenov found that in the
free-energetically favored brush configuration chain ends are in fact
distributed throughout the brush, i.e.  the ``Alexander-de Gennes''
picture in which all ends lie at height $h$ is wrong.
Correspondingly, the pressure field is non-uniform and quadratic,
$P(z)= C P_0 (1-z^2/h^2)$ where $C=3 \pi^2 / 4$.

Extending Semenov's analysis, we minimize $F$ with respect to both
polymer plus nanoparticle fields.  At moderate volume fractions
$F_{nano} \simeq \int d{\bf r} (\phi_{nano}/b^3) \ln
(\phi_{nano}/e)$ is dominated by translational entropy and we find in
equilibrium
\begin{eq}{filthy-toilet}
\phi_{nano}(z) = \phi_{nano}(h) \ e^{-P(z) b^3}
\end{eq}
where the pressure field remains quadratic but the substrate pressure
and brush height are increased. Noting the pressure vanishes at the
brush surface $z=h$, and is approximately linear for small depths, we
write $\phi_{nano}(z) \twid\ e^{-(h-z)/\delta}$ where $\delta \equiv
h(b^*/b)^3$ and $b^* \approx (N/h)^{2/3} = 1/\sigma^{2/3}$.  Thus for
$b\gg b^*$ the penetration depth $\delta\twid 1/b^3$ is much less than
the full brush height. It decreases with increasing particle size
until $\delta =b$ defines the maximum sized particle $b=\bmax \approx
(N/\sigma)^{1/4}$ which can enter the brush.  In the partial
penetration regime, $b^*<b<\bmax$, the maximum particle volume
fraction is $\phimax = \delta/h \twid 1/b^3$.  Above this level,
excess nanopartcles are expelled outside \citeben{nonwet}.

\begin{strip_ft}
{0}
{0}{7}{figure3}
{0}
{0}{7}{\caption[Propagation]{
\renewcommand{\baselinestretch}{1} \footnotesize \bf
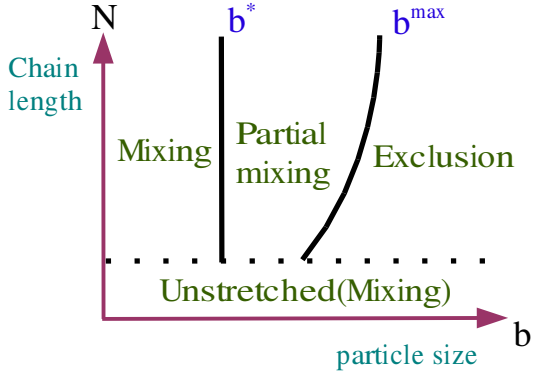
Polymer-soluble nanoparticles (case (A)): phase diagram as a function
of particle size (b) and polymer chain length (N). Particles smaller
than $b^*$ mix freely in the film while those larger than $b_{max}
\sim N^{1/4}$ are excluded. In the intermediate regime, $b^* < b <
b_{max}$, partial film penetration occurs to a depth $\delta \sim
1/b^3$ with loading capacity $\phi_{max} = \delta/h \sim 1/b^3$. 
}
\label{soluble_phase_diagram}}
\end{strip_ft}

These results are represented in fig. \ref{soluble_phase_diagram}.
Their physical origin is clear.  Because of Semenov's chain end
relaxation effect, the edge of the brush is much softer than elsewhere
and the pressure profile increases monotonically as one moves away
from the free brush surface.  This provides a ``buoyancy'' force on
any inclusion, tending to push particles to the brush surface.
Mixing entropy opposes this, and the balance leads to the density
field of eq. \eqref{filthy-toilet}.  The penetration depth $\delta$ is
obtained by balancing the entropy, of order $kT$, with $P(\delta) b^3$
which is the particle's chain stretching energy penalty at depth
$\delta$.

Let us now turn to case (B), nanoparticles insoluble in the polymer
medium, i.e. the nanoparticle-polymer interfacial tension
$\gamma_{np}$(see below) exceeds a critical (positive) value.
After being forced into the brush, van der Waals attractions coalesce
them into increasingly big aggregates \citeben{liu:nanoletters_prl}.
Unchecked, phase separation would result.  Now to understand how the
brush may intervene, we must go beyond the analysis above which
produced density profiles homogeneous in the lateral directions
($x,y$).  This is meaningful only for nanoinclusions which can be
treated as fluid-like perturbations, i.e. those smaller than the brush
``blob'' size $\xi$.  This is the scale over which a brush chain's
statistics are of unperturbed Gaussian character, while on larger
scales the global stretching is felt \citeben{gennes:book}.  When
large aggregates grow $(b>\xi)$ the brush is strongly disrupted
locally and (x,y) symmetry is broken.  From a theoretical standpoint,
the problem is intrinsically non-perturbative.  Amongst previous works
addressing this problem
\citeben{solistang:perturb_prl,fredrickson:surfacemode_prl}, an important
contribution was due to Williams and Pincus
\citeben{williamspincus:hydroanal_prl} who proved that an object in a
strongly stretched brush is formally analogous to irrotational steady
state inviscid flow past the same object. Polymer configurations map
to flow streamlines, and the brush energy equals the fluid kinetic
energy \citeben{williamspincus:hydroanal_prl}.

Exploiting this analogy, by using flow solutions past various objects
\citeben{lamb:hydro} we have calculated the brush energy cost $\Delta
F$ for nanoparticles to form aggregates of different sizes and
shapes. (see fig. \ref{aggregate_shape})
We find that momentum conservation in the hydrodynamic analogy
translates to the statement that for a given aggregate volume $\Omega$
there is a baseline free energy cost of exactly $2\Po \Omega$.
Additional free energy cost depends on aggregate shape: we find
$\Delta F= (5/2)\Po\Omega$ and $3\Po\Omega$ for, respectively, spheres
and for long cylinders parallel to the substrate.  Most interestingly,
{\em disk-shaped} aggregates of diameter $l$ and thickness $d$ are
much more costly: $\Delta F= (4/3\pi) \Po\Omega (l/d)$ is dominated by
the aspect ratio factor, $l/d\gg 1$.

While helpful, these results are not quite conclusive since the
hydrodyanmic analogy applies only to the simplified ``Alexander-de
Gennes'' brush \citeben{williamspincus:hydroanal_prl}.  How does chain
end-annealing modify these findings?  To answer this, we have
established rigorous bounds on $\Delta F$ for an inclusion at height
$z$ in the true end-annealed brush: (i) We find a lower bound of $P(z)
\Omega$, closely related to the momentum conservation condition, and
corresponding to a fluidized aggregate laterally smeared to infinity;
(ii) a shape-dependent upper bound results from a constructed brush
configuration correctly respecting incompressibility.  For the sphere
and cylinder cases the bounds differ only in the prefactor.  For the
disk, we find an upper bound $\Delta F < (2/3\pi) \, P(z) \Omega
\,(l/d)$ which equals our result in the Alexander-de Gennes picture
apart from the prefactor. Thus (see fig. \ref{aggregate_shape})
                                        \begin{eq}{shit-smeared-on-face}
\Delta F =  \alpha\, P(z) \,\Omega\ \ (\mbox{sphere})\comma\ \
    \Delta F = \beta\, P(z) \, \Omega\ \ (\mbox{cylinder}) \comma\ \
        \Delta F = \gamma \, P(z)\, \Omega\ {l \over d}\ (\mbox{disk})   \comma   
                    \end{eq}
where $\alpha, \beta$ are constants and $1<\alpha<5/4$,
$1<\beta<3/2$. For the disk we conjecture $\gamma <2/3\pi$ is a
constant.

We conclude that the effect of end annealing is to replace $\Po$ with
the pressure $P(z)$ at the aggregate location, and to modify
prefactors.  Eq. \eqref{shit-smeared-on-face} tells us that as
nanoparticle aggregates grow their elongation in one lateral direction
(cylindrical shape) invokes little penalty, whereas the brush strongly
opposes simultaneous growth in two lateral directions (disk-like) as
shown in fig. \ref{aggregate_shape}.  Macroscopic phase separation
within the film is thus suppressed.

\begin{strip_ft}
{0}
{0}{7}{figure4}
{0}
{0}{7}{\caption[Propagation]{
\renewcommand{\baselinestretch}{1} \footnotesize \bf
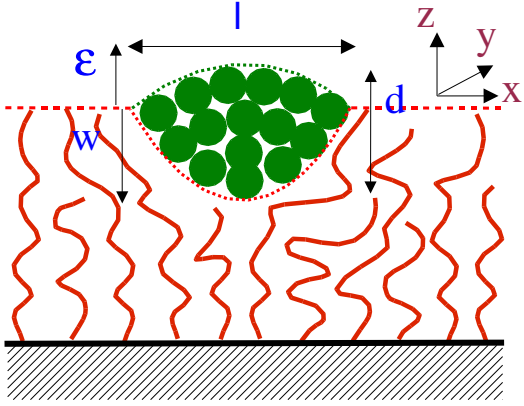
To minimize its free energy a nanoparticle
aggregate at the polymer film-air interface chooses a baguette shape
of length $L$ in the $y$ direction(perpendicular to the plain of the
paper). Its cross section comprises 2 half lenses above and below the
film surface.

}
\label{nanolens}}
\end{strip_ft}

However, due to the $P(z)$ factors in eq. \eqref{shit-smeared-on-face}
a ``buoyancy'' force, strengthening as aggregates grow, thrusts them
towards the polymer-air interface.  The final state thus hinges on the
3 relevant surface tensions: $\gamma_{np}$ (nanoaggregate-polymer),
$\gamma_{na}$ (nanoaggregate-air) and $\gamma_{pa}$ (polymer-air).
These determine the familiar {\em spreading coefficient}
\citeben{aveyardclint:lens_prl} $S\equiv
\gamma_{pa}-\gamma_{np}-\gamma_{na}$ and {\em entry coefficient}
\citeben{aveyardclint:lens_prl}
$E\equiv\gamma_{pa}+\gamma_{np}-\gamma_{na}$.  The free energy of a
nanoparticle aggregate at the polymer-air interface reads
                                                \begin{eq}{lens}
\Delta F = \frac{P_0  L l^3 w^2}{h^3} \,  f_1(L/l)
	-   S \, L l  \, f_2(L/l)
   +  \square{
\gamma_{na} {L \epsilon^2 \over l} 
   +  \gamma_{np} {L   w^2      \over l}
} \, f_3(L/l) \period
                                                                \end{eq}
Fig. \ref{nanolens} depicts the aggregate of length $L$ (y
direction) with lens-shaped cross-section of width $l$ (x direction)
and bulging a depth $w$ into the brush and $\epsilon$ into the air
($\epsilon+w=d$).  The baguette shape generalizes to anisotropic
viscoelastic media the well known result that a droplet on a liquid
surface assumes a circular lens shape to minimize its free energy
\citeben{aveyardclint:lens_prl}.  The first term is the brush stretching
energy penalty for a surface depression of depth $w$, width $l$ and
length $L$ \citeben{fredrickson:surfacemode_prl}, the second results from
the baguette displacing an area $Ll$ of air-polymer interface and the
last 2 from extra surface area created by each half lens.  The factors
$f_i\equiv \alpha_i + \beta_i l/L$ add extra areas due to the
baguette's end-caps (located at $y=0, L$) where $\alpha_i, \beta_i$
are constant geometric factors of order unity.

This free energy is somewhat simplified for brevity's sake (e.g. the
baguette may random walk and branch in the x-y plane) but retains the
essential features.  Next we minimize the free energy per unit volume
for an ensemble of aggregates, $\Delta F/\Omega$ (where $\Omega = Lld
(A + Bl/L)$ with $A,B$ constants), subject to the constraint $\epsilon
< \epsilon_{\small max}$ where $\epsilon_{\small max}$ is the maximum
droplet growth into the air medium allowed by the experimental setup
or kinetics (as discussed previously).  We find: (i) If $S>0$
(i.e. nanoparticles {\em wet} the polymer material) an infinite
nanoparticle monolayer spreads onto the polymer film surface,
$L,l\gt\infty,\ d\gt 0$.  Physically, there is no advantage in
penetrating the brush because droplet entry into the polymer-air
interface is favored since $E>0$ (since $E>S$ when $\gamma_{np}>0$).
(ii) If $S<0$ and $E>0$, we find $l \approx \delta^{1/4} h$ where
$\delta \equiv \gamma_{np}/P_0 h$ is typically of order unity, $d$ is
also of order $h$ and $L\rightarrow \infty$. That is, nanoaggregates
grow freely in one lateral direction, but the brush stabilizes growth
in the other lateral direction, selecting a characteristic width.  (We
find a similar stabilization in the slightly more complex case of
$E<0$.)

\ignore{
Again, cylindrical-like growth is favored over disk-like growth: this
is the direct extension of our previous analysis for intrabrush
aggregates.  The similarity of the conclusions is very natural because
the physical origin of this anisotropic shape selection is the energy
penalty associated with stretching of brush chains.
} 

Let us now compare to experiment.  In
ref. \citenum{liu:nanoletters_prl}, TEM images of $2.5nm$
PEP-insoluble (case (B)) dodecanethiol-stabilized gold nanoparticles
in PEP thin film brushes show nanoparticles apparently macroscopically
phase separated into film surface monolayers, suggesting $S>0$.
However, $5nm$ films exhibited elongated aggregates with a lateral
scale of a few $nm$, a brush-selected morphology similar to that
predicted here.  These latter are actually collapsed brushes
\citeben{liu:nanoletters_prl}, a case not considered here.  We
speculate that the thinness of these films prevents wetting and shifts
the behavior into the size-stablized regime described above.  Thomas
and coworkers \citeben{fogg:cluster_prl} report $4.5nm$ ZnS-coated
CdSe particles insoluble (case (B)) in NBPBD-NBP block copolymer films
of thickness $30-50nm$ spreading macroscopically on the film surface
(fig. 5 of ref. \citenum{fogg:cluster_prl}). This suggest $S>0$
(though loss of the capping species may explain this).  But when the
solvent was evaporated rapidly, aggregates were instead elongated with
finite lateral scale (fig. 6 of ref. \citenum{fogg:cluster_prl}).
This is consistent with a snapshot of the shape-directed aggregate
growth kinetics described here.  Lauter-Pasyuk et
al. \citeben{hamdoun:composite_prl} studied $Fe_2 O_3$ nanoparticles
in lamallar PS-PBMA diblock materials.  A corona of PS oligomers
apparently rendered particles soluble (case (A)) in the PS domains.
Neutron specular reflection data suggested $6nm$ nanoparticles remain
near the PS center plane whilst $4nm$ particles penetrate further.
Tentatively, we estimate $3.8\lsim b^* \lsim 6.5nm$, suggesting the
partial mixing regime of fig. \ref{soluble_phase_diagram} with the
penetration depth $\delta\twid 1/b^3$ larger for the smaller
particles.  Balazs and coworkers \citeben{balazs:numeric_prl}
numerically solved equations describing lamellar diblocks with
nanoparticles soluble in one polymer species (case (A)).  We estimate
$(b/b^*)$ values of 2 and 1.3, indicating the partial mixing regime
(see fig. \ref{soluble_phase_diagram}).  Consistent with this,
nanoparticle density profiles peaked at and decayed away from the
lamellar center plane.

In summary, ultrathin films of end-tethered polymers can actively
organize nanoparticles to produce novel composite materials. The
polymer medium generates effective anisotropic interactions between
nanoparticles which direct and stabilize the growth of aggregates
within the film and determine the ultimate morphology of the
nanoparticle dispersion. The origin of these effects is that the
stretching energy cost for polymer brush chains depends strongly on
aggregate shape. Our theory suggests two alternative strategies for
experimentalists aiming to mix nanoparticles into polymer brushes or
lamellar phases. (1) Compatibilize particles to be soluble in the
polymer medium and synthesize them as small as
possible; particles above a certain size ($b>b_{max}$) will simply not
remain within the film. For example, with $h/R_g = 2$ where $R_g =
N^{1/2}$ is the unperturbed polymer coil size
\citeben{liu:nanoletters_prl} one has $b_{max} \approx R_g /3 \approx 3
nm$ for monomer size $3nm$ and $N=1000$. (2) To stabilize giant
elongated aggregates embedded in the film but near its surface,
nanoparticles should be insoluble in the polymer and {\em non-wetting}
on the polymer surface ($S<0$).  Thus, as important as the
nanoparticle design is the design of the polymer medium whose
incompatibility with air(or whichever is the third medium) should be
minimized. Ideally, the polymer should be less incompatible with air
than the nanoparticles are ($\gamma_{pa} < \gamma_{na}$): this guarantees a
stable morphology.

This work was supported by the MRSEC Program of the NSF, award
no.~DMR-98-09687.  We thank Chris Durning, Rasti Levicky, Royce Murray
and Qingbo Yang for stimulating discussions.


\pagebreak

\bibliography{nanoprl}

\begin{thebibliography}{10}
\bibitem{nonwet}
In the soluble case, for simplicity we assumed particles are non-wetting on the
  brush surface ($S<0$).
\end{thebibliography}

\end{document}